\documentclass[conference]{IEEEtran}
\IEEEoverridecommandlockouts

\usepackage[T1]{fontenc}
\usepackage[utf8]{inputenc}
\usepackage{inputenc}
\usepackage[english]{babel}
\usepackage{ulem}
\usepackage{comment}

\usepackage[table]{xcolor}

\usepackage{color}

\usepackage{algorithm,algpseudocode} 

\usepackage{hyperref} 
\usepackage{multirow}

\usepackage{amsmath,amssymb,amsfonts,amsthm}
\usepackage{bbding}
\usepackage{tikz}
\usepackage{csquotes}
\usepackage{dashbox}
\usepackage{todonotes}
\usepackage{enumitem}   

\usepackage{graphicx,subfigure}
\definecolor{bblue}{HTML}{4F81BD}
\definecolor{rred}{HTML}{C0504D}
\definecolor{ggreen}{HTML}{9BBB59}
\definecolor{ppurple}{HTML}{9F4C7C}
\usepackage{xcolor}
\usepackage{tikz}
\usepackage{pgfplots}

\definecolor{findOptimalPartition}{HTML}{D7191C}
\definecolor{storeClusterComponent}{HTML}{FDAE61}
\definecolor{dbscan}{HTML}{ABDDA4}
\definecolor{constructCluster}{HTML}{2B83BA}

\usepackage{listings}
\usepackage{trace}
\usepackage{color}
\usepackage{makeidx}
\usepackage{mdframed}

\usepackage{listings}
\usepackage{color}
\usepackage{footmisc} 

\usepackage{fancyhdr} 
\usepackage{graphicx}

\newcommand{\qw}[1]{\textcolor{blue}{qw: #1}}

\usepackage{indentfirst}

\usepackage{booktabs}    

\usepackage{dashbox}
\usepackage{tikz}

\usepackage{ulem}

\usepackage{threeparttable}

\usepackage{url}

\usepackage[
	n,
	operators,
	advantage,
	sets,
	adversary,
	landau,
	probability,
	notions,	
	logic,
	ff,
	mm,
	primitives,
	events,
	complexity,
	asymptotics,
	keys]{cryptocode}

\usepackage{listings}
\usepackage{color}

\definecolor{dkgreen}{rgb}{0,0.6,0}
\definecolor{gray}{rgb}{0.5,0.5,0.5}
\definecolor{mauve}{rgb}{0.58,0,0.82}
\definecolor{backcolour}{rgb}{0.95,0.95,0.92}
\definecolor{mycolor}{rgb}{1,0.99,0.7}

\begin{document}

\title{Frontrunning Block Attack in PoA Clique: \\ A Case Study\thanks{This work was in part presented at IEEE ICBC 2022.}}


\author{\IEEEauthorblockN{Xinrui Zhang \IEEEauthorrefmark{1},\thanks{$\eth$: Corresponding author.}
Qin Wang \IEEEauthorrefmark{3},
Rujia Li \IEEEauthorrefmark{1}\IEEEauthorrefmark{2},
Qi Wang \IEEEauthorrefmark{1}$^\eth$
}
\IEEEauthorrefmark{1} {\textit{Southern University of Science and Technology}}, China.\\ 
\IEEEauthorblockA{\IEEEauthorrefmark{2} {\textit{University of Birmingham}}, United Kingdom.\\ 
\IEEEauthorrefmark{3} {\textit{CSIRO Data61}},  Australia. \\ 
}
}


\maketitle

\IEEEpubidadjcol

\begin{abstract} 
As a fundamental technology of decentralized finance (DeFi), blockchain's ability to maintain a distributed fair ledger is threatened by manipulation of block/transaction order. In this paper, we propose a frontrunning block attack against the Clique-based Proof of Authority (PoA) algorithms. Our attack can frontrun blocks from honest in-turn sealers by breaking the proper order of leader selection. By falsifying the priority parameters (both \textit{difficulty} and \textit{delay time}), a malicious out-of-turn sealer can always successfully occupy the leader position and produce advantageous blocks that may contain profitable transactions. As a typical instance, we apply our attack to a mature Clique-engined project, HPB (\$3,058,901, as of 
April 2022). Experimental results demonstrate the effectiveness and feasibility. Then, we further recommend fixes that make identity checks effective. Our investigation and suggestion have been submitted to its official team and got their approval. We believe this work can act as, at least, a warning case for Clique variants to avoid repeating these design mistakes. 

\end{abstract}

\section{Introduction}

Proof of Authority (PoA) consensus algorithm was first proposed by Wood \cite{ethereum-poa}\cite{szilagyi2017eip} and further used in Geth~\cite{geth} and Parity~\cite{parity}. As an implementation of PoA algorithms, Clique has become a mainstream design since it is merged into the Ethereum mainnet~\cite{samuel2021choice}. Clique relies on a limited number of authorities (denoted by \textit{sealers}) and their reputation to achieve consensus. Instead of selecting one mining leader in each round, Clique allows multiple sealers to produce blocks in parallel with \underline{different} priorities. This mechanism makes Clique only process one round of messages across multiple sealers. Fewer message exchanges result in faster confirmation and better performance. Owing to those advantages, Clique has recently emerged as one of the most prevailing consensus algorithms and inspires many variants that have been applied to well-known projects (cf. Table~\ref{tab-projects}), such as GoChain~\cite{gochain}, Binance chain~\cite{binance}, POA network \cite{poa2021} and ConsenSys \cite{consensys}. 

\begin{figure}[h!]
    \centering
    \includegraphics[width=\linewidth]{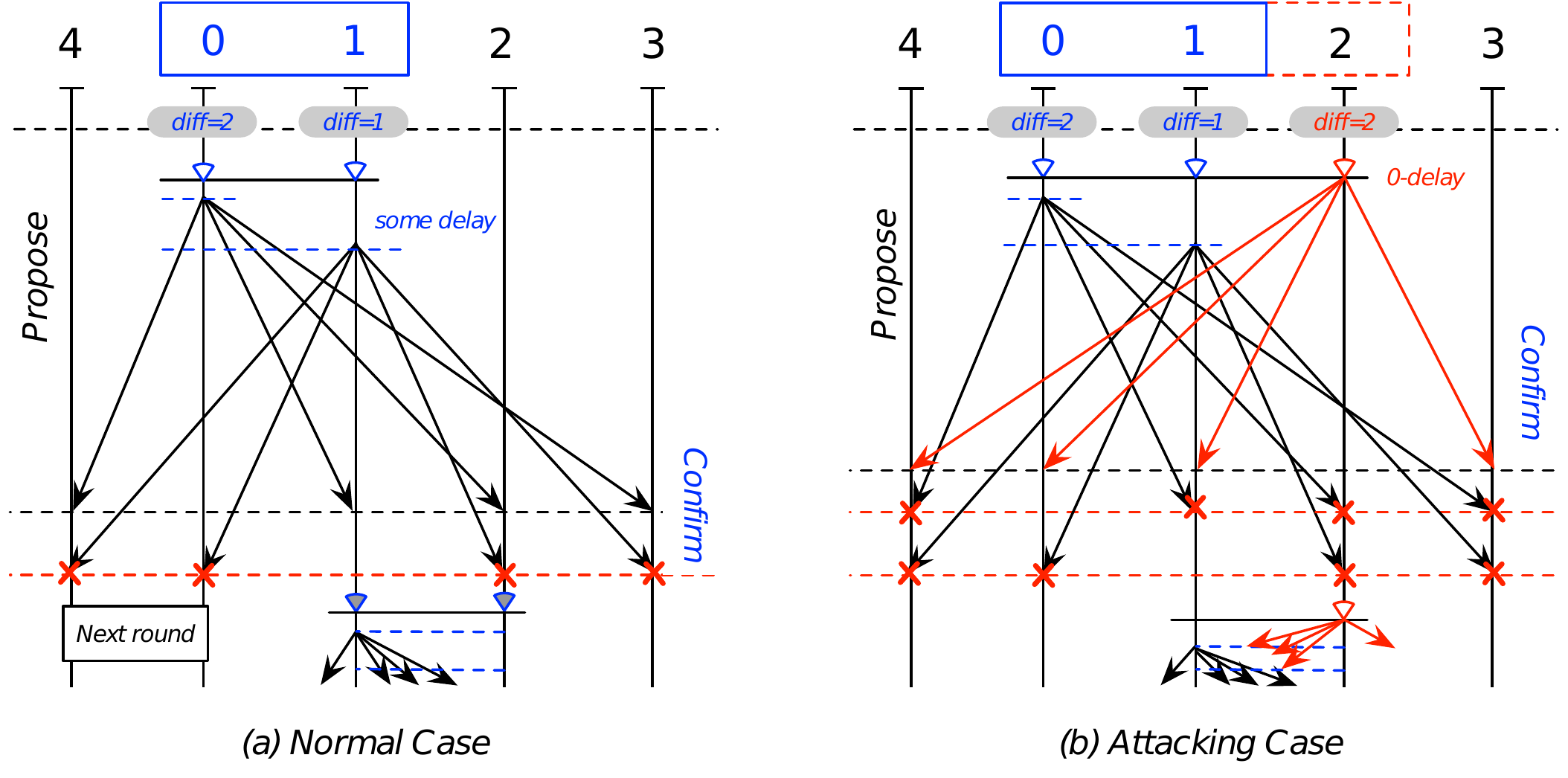}
   \caption{Operation Mechanisms of Clique and Our Attack}
   \label{fig-poa}
\end{figure}

In the original design of Clique, a set of sealers are allowed to propose blocks in the same epoch, where an in-turn sealer (equiv. leader) proposes the first-priority block, while edge-turn sealers create second-priority blocks with a random delay (cf. Figure \ref{fig-poa}.(a)). Since more than one blocks are created, forks may occur. Clique adopts a simplified version of GHOST protocol \cite{yellowpaper}\cite{sompolinsky2015secure} to select branches. A higher priority ensures that a block is more likely to be included in the final chain. 
Under the default assumption of Clique, the majority of sealers are honest. An in-turn sealer is elected each round in a round-robin way. 
This is inherently determined by a predefined formula combining the block index and the number of sealers. For more details, see example~\cite{wang2022poa}\cite{de2018pbft}. 
However, in real-world settings, a malicious sealer may locally break the honest order of sealer rotation to occupy the leader position and thus create profitable blocks by imitating the \textit{leader's priority}. This attack may indirectly lead to Miner Extractable Value (MEV) issues \cite{daian2020flash}\cite{torres2021frontrunner} and cause considerable financial loss to clients. Fortunately, Clique mitigates this issue by adding a verification mechanism to check the identity of each miner. However, several variants of Clique are still vulnerable to the risk of frontrunning by malicious non-in-turn\footnote{Non-in-turn sealers contain both edge-turn (produce secondary-priority blocks, \textit{diff=1}) and out-of-turn sealers (cannot produce blocks,  \textit{diff=0}) \cite{wang2022poa}.} sealers and can therefore experience a block-level frontrunning attack.

In this paper, we dive into Clique-based projects, to explore their improper ways of using Clique, and identify the overlooked vulnerabilities that may seriously threaten the consensus stability. We launch an adversarial frontrunning attack towards its basic mechanism of sealer rotation to occupy the leader's position. To this end, we apply our attack to a mature \textit{in the wild} project High Performance Blockchain (HPB)~\cite{hpb-poa} under an isolated environment due to ethical principles. HPB is designed as a permissioned blockchain applying PoA with dynamic committee to seal blocks. 
A sealer is assumed to be malicious and can falsify priority parameters of \textit{difficulty} (\textit{diff} for short) and \textit{delay time}. The simulation results show that the frontrunning blocks created by this malicious sealer are always accepted by peer sealers due to its falsified \textit{diff}(=2) and \textit{0-delay}ed block production time (cf. Figure \ref{fig-poa}.(b)).  We then provide our intuitive but useful suggestion to its official team. In a nutshell, the contributions are as follows.

  \begin{itemize}
    \item[-] We proposed a novel type of attack against PoA protocols, especially for Clique-based implementations. The proposed attack can frontrun the leader's position by falsifying priority parameters (\textit{diff} and \textit{delay time}) to produce blocks containing profitable transactions. 
    
    \item[-] We investigated the assumptions and properties of Clique and dived into the root reason that causes such attacks. 
    
    \item[-] We applied our attack to a mature Clique variant (HPB) as a case study. The results indicate that our attack strategies can cause actual damages.
    
    \item[-] We provided recommendations on how to fix the system against the proposed attack. We further submitted the issue to its official team and obtained their approval~\cite{go-hpb-pr}.
 
 \end{itemize}

\smallskip
\noindent\textbf{Paper Structure.} Section~\ref{sec-clique} outlines the general construction of Clique. Section~\ref{sec-attack} introduces the details of the frontrunning block attack. Both experimental simulation and evaluation are presented in section~\ref{sec-experiment}. Section~\ref{sec-seurity} provides the corresponding security analysis. Countermeasures are proposed in section~\ref{sec-countermeasures}. Finally, section~\ref{sec-rw} exploits related studies, and section~\ref{sec-conclusion} concludes our work.

\section{Clique Model}
\label{sec-clique}

In this section, we introduce the Clique model by separately presenting the threat model, underlying assumptions and the detailed algorithm.

\smallskip
\noindent\textbf{Threat Model.} Designed for permissioned blockchain systems, Clique only allows a limited number of ($N$) sealers to participate in the consensus procedure~\cite{samuel2021choice}. The network is assumed to be partially synchronized, where all messages will be eventually delivered, but the delay time is unknown. Meanwhile, the system requires a majority of sealers to be honest: any sealer may become malicious and do evil activities such as dropping, delaying, duplicating, or even faking messages, but most sealers are honest. Our attack follows the above settings with an additional assumption: \textit{we allow an adversarial sealer to be profitable and to obtain extra interests by frontrunning the leader position to produce blocks}. We emphasize that our attack strategy does not break security assumptions because the assumed malicious adversary obeys all the principles defined in the source code. Instead of \textit{consistency} and \textit{liveness} \cite{garay2015bitcoin}, we target at the property of \textit{fairness}~\cite{kelkar2020order}\cite{kelkar2021order} that focuses on fair transaction ordering.

\smallskip
\noindent\textbf{Algorithm Description.} Clique allows multiple sealers to  propose blocks in parallel. Particularly, in each round, an in-turn sealer will be elected as the leader according to the block index and the total number of sealers (cf. line~\ref{clique:line:leader} Algorithm \ref{alg-proposeclique}). The election happens in a round-robin way. Certain sealers whose numbers follow the in-turn sealer's become edge-turn sealers. The amount of out-of-turn sealers is set to the \textit{smallest majority}, which is $\left\lfloor\frac{N}{2}\right\rfloor+1$. By contrast, $\left\lceil\frac{N}{2}\right\rceil-2$ sealers following the in-turn sealer are labeled as edge-turn sealers. In-turn sealer and edge-turn sealers compose the \textit{largest minority}.
Both the in-turn sealer and edge-turn sealers are allowed to propose blocks. Besides \textit{diff}, the main difference between them is that the edge-turn sealers (\textit{diff}=1, equiv. \textit{b.weight}=1, line~\ref{clique:line:weight1}) can only sign a new block after the delay (line \ref{clique:line:sleep}), whereas the in-turn sealer (\textit{b.weight}=2, line~\ref{clique:line:weight2}) signs and broadcasts the new block instantly (still with a short delay). This parallel block-producing mechanism effectively prevents a single Byzantine sealer from causing havoc on the entire network. However, forks may 
occur as the system needs time to decide which block is valid. \textit{Clique} solves this issue by using a simplified version of GHOST protocol~\cite{yellowpaper}: in-turn sealers' blocks will be granted with a higher weight to ensure that they will be finally confirmed by a majority of sealers. In contrast, a random delay with a \textit{wiggle}  (line~\ref{clique:line:delay}) slows the block generation of an edge-turn block-producing sealer. In this way, an in-turn sealer's block has a better chance of being accepted by peers than edge-turn sealers.

\begin{algorithm}
\caption{Block Propose and Block Verify in Clique}
\label{alg-proposeclique}
\begin{algorithmic}[1]
\Procedure{Propose}{$sealer_i$} \Comment{\textcolor{blue}{propose a
block}} 
\While{$(\mathsf{true})$} 
\State $n \gets lastblock.number$
\State $delay \gets block.time - now$
\State $N \gets |sealers|$
\State $\textbf{wait until} \neg sign\_recently(sealer_i,n)$ \label{clique:line:recent}
\Comment{\textcolor{red}{attack}}
\State $\textbf{if}\, ((n+1) \mod N) = i$ \textbf{then} \label{clique:line:leader}
\State $\qquad weight = 2$ \Comment{\textcolor{blue}{in-turn sealing}}
\label{clique:line:weight2}  
\State $\textbf{else}$
\State $\qquad weight = 1$
\Comment{\textcolor{red}{attack}}
\label{clique:line:weight1}
\State $\qquad delay = rand(delay + (\frac{N}{2}+1) * 500 ms)$
\label{clique:line:delay}
\State $sleep(delay)$ \label{clique:line:sleep}
\Comment{\textcolor{red}{attack}}
\State $block \gets sign(TXs, weight)$ \Comment{\textcolor{blue}{seal a block}}
\State $broadcast(block)$ \Comment{\textcolor{blue}{send the block}}
\EndWhile\label{euclidendwhile}

\EndProcedure
\State $-----------------------------$
\Procedure{Verify}{$sealer_i$}
\Comment{\textcolor{blue}{verify a
block}}
\State $\neg sign\_recently(sealer_i,block.number)$ \label{clique:line:verrecent}
\State $\neg difficulty\_check(sealer_i,block.weight)$ \label{clique:line:verdiff}
\State $\neg inturn\_check(sealer_i,block.number)$ \label{clique:line:verinturn}
\State $ \textbf{return}$ $\mathsf{true}$ \Comment{\textcolor{blue}{block is valid}}
\EndProcedure

\end{algorithmic}
\end{algorithm}

Verification procedure is evoked when a sealer node receives a block. According to the original design of Clique, three core verification functions are deployed. Block receivers first check whether the block's proposer has recently signed a block. Sealers maintain a local list, and reject blocks whose proposer is labeled as \textit{recently signed}. Secondly, the value of block \textit{diff} will be checked to ensure that it is restricted to $1$ or $2$. Finally, receivers should have an additional check on blocks with a difficulty $2$, to ensure that the proposer is a valid leader.

\begin{figure*}[ht]
    \centering
    \includegraphics[width=1\linewidth]{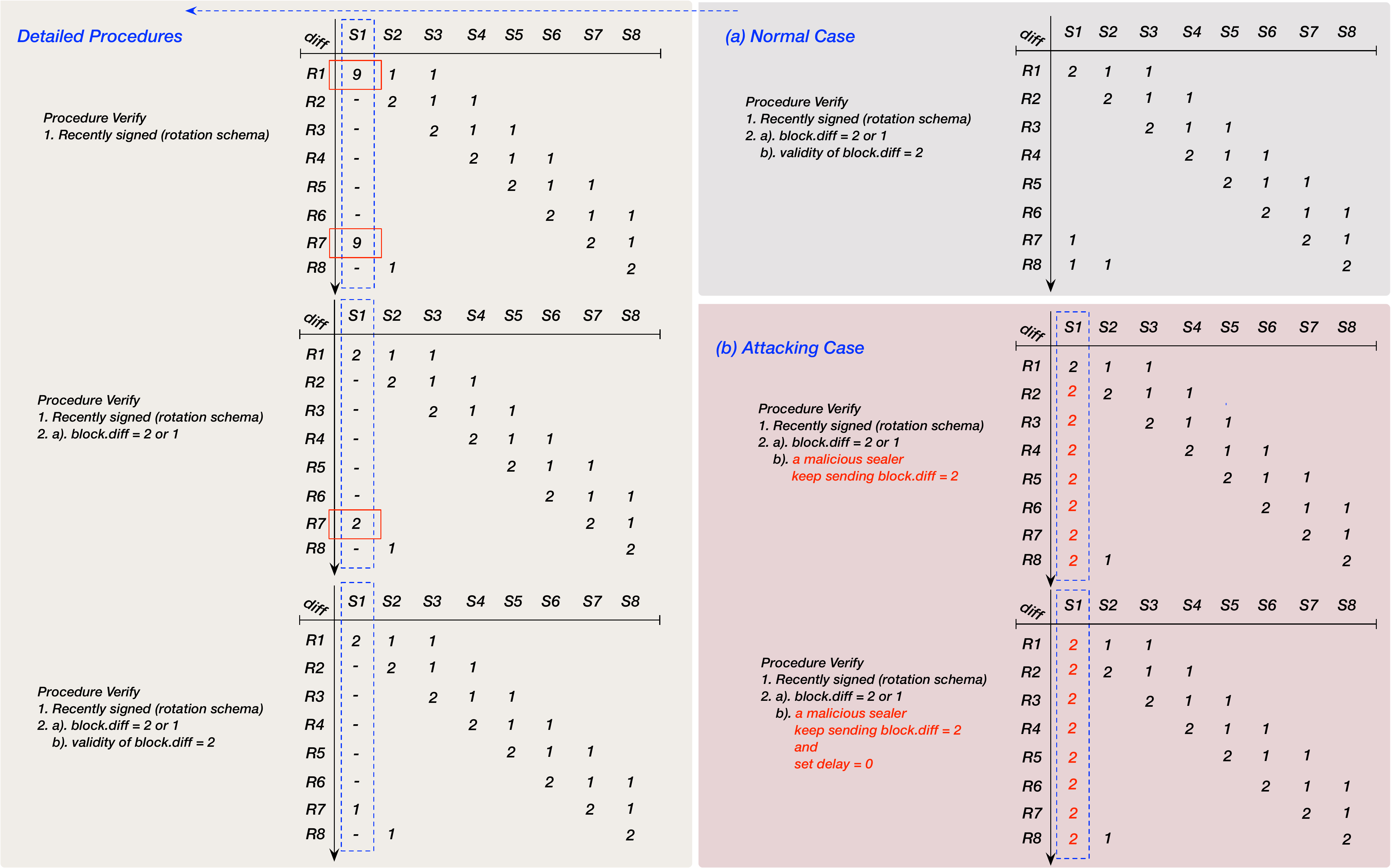}
   \caption{Detailed Procedures of Verification and Our Attack}
   \label{fig-attack}
\end{figure*}

\section{Frontrunning Block Attack}
\label{sec-attack}

Recently, with the development of decentralized finance (DeFi)~\cite{zetzsche2020decentralized} and non-fungible token (NFT)~\cite{Wang2021NonFungibleT}, a frontrunning attack have come to the fore. In a blockchain system, rushing adversaries~\cite{kosba2016hawk} may discard/replace/reorder honest transactions for arbitrage. 
For instance, in the application of decentralized exchanges, an adversary can place a transaction before others' if the price of an asset is going to increase~\cite{zhou2021high}. This attack usually happens in the transaction pending stage when transactions flow from the local mempool to a block~\cite{eskandari2019sok}. 
Another typical instance is sandwich attack~\cite{zhou2021high}. 
Instead of exploring the transaction-level attack, we focus on our block-level frontrunning attack. Block-level means that the block should be proposed in a proper turn. This requires their producers (sealers) to faithfully take turns to mine blocks without breaking the prearranged turn.

As discussed, the Clique algorithm mainly uses the rotation schema to select a leader and the priority parameters (\textit{diff} and \textit{delay time}) to increase acceptance rate of the leader's block. As an example shown in Figure~\ref{fig-attack}, eight sealers ($S_1$ -- $S_8$) run the consensus algorithm. Among them, $S_1$ is supposed to be a malicious sealer while the other seven sealers are honest. An ideal frontrunning block attack aims to disrupt the rotation schema and priority parameter assignment: (i) \textit{Sealer}$_1$  can keep sealing blocks, ignoring the rotation; (ii) \textit{Sealer}$_1$  can tamper the proposed block weight (\textit{diff}) from $1$ to $2$; (iii) or an arbitrary invalid value, e.g., 9. Such an attack is deemed as successful if \textit{Sealer}$_1$  can seal blocks out of the legal turn among peer blocks; meanwhile, all blocks of \textit{Sealer}$_1$ should pass the verification and finally get accepted by honest sealers.

\section{Attacking Implementation}
\label{sec-experiment}
This section outlines a series of attacks on commercial blockchain networks using Clique-based consensus algorithms. Our target is to achieve an ideal attack by disrupting the rotation schema and priority parameter assignment. As an example of a successful attack, we depict an attacking simulation on a widely used blockchain, HPB. Notably, we emphasize that HPB is merely one case, and our target is to help communities to raise awareness for design pitfalls, especially for subtle logic errors hidden in complex code.

\smallskip
\noindent\textbf{Experiment Configuration.}
We establish the HPB private blockchain on a local workstation with Centos 7 x86\_64 operating system. We simulate multiple consensus sealers by using different ports. Our private blockchain consists of six nodes: five of them are sealer nodes ($N\in \{0,1,...,4\}$) responsible for mining, and the other is a bootnode~\cite{private-net} for peer detection. We set the block generation rate at five seconds, and then continuously send transactions to any sealer at a constant rate ($10$ TXs/s) in a fixed period ($30$ minutes).

\smallskip
\noindent\textbf{Experimental Procedure.} 
The experiment is conducted from two sides. Firstly, we consider a normal case where the operations of all six nodes are honest. We compile the HPB client \textit{go-hpb}  \cite{go-hpb} with official source code and refer to the original program as $\mathsf{P_{\mathcal{O}}}$. Then, we faithfully execute $\mathsf{P_{\mathcal{O}}}$. Five enrolled sealers are assumed to be honest and will follow the rule of consensus honestly. The results are measured by the output address \textit{addr} of the in-turn sealer in each round of block production (See code in Listing~\ref{lst:label}). These experiments are used as a benchmark against the experiments suffering attacks.

With the intention of keeping proposing blocks, malicious nodes need to modify the source code of \textit{go-hpb} and produce  $\mathsf{P_{\mathcal{M}}}$ where local block generation constraints (cf. line~\ref{clique:line:recent} Algorithm~\ref{alg-proposeclique}) are nullified, and block \textit{diff} is fixed at $2$ (line~\ref{clique:line:weight1}). On the code level, simply omitting code related to those constraints may achieve the nullification and breaks the rotation schema. Falsifying the difficulty assignment process by changing \textit{diff=1} to \textit{diff=2} will set \textit{diff} to a static $2$. Moreover, omitting the \textit{delay time} related code (cf. line~\ref{clique:line:sleep}) gives malicious node further priority in block confirmation.

Therefore, secondly, we simulate the frontrunning attack, in which one (denoted by \textit{Sealer}$_2$) of the five sealers is malicious with intentions to frontrun the position of an in-turn sealer, while the other four sealers (\textit{Sealer}$_{\{0/1/3/4\}}$) remain honest and execute  $\mathsf{P_{\mathcal{O}}}$ faithfully. \textit{Sealer}$_2$ first modifies the source code, \textit{falsifies its \textit{diff} to a static 2} (cf. line~\ref{clique:line:weight1} Algorithm \ref{alg-proposeclique}) and \textit{omits its delay time} (cf. line~\ref{clique:line:sleep} Algorithm \ref{alg-proposeclique}), so that \textit{Sealer}$_2$ has the same acceptance weight (first-priority) as the honest in-turn sealer with even faster confirmation. The condition of \textit{sign\_recently} (cf. line~\ref{clique:line:recent} Algorithm \ref{alg-proposeclique}) is also omitted locally, so that \textit{Sealer}$_2$ can produce blocks continuously even if it has signed a block recently.
\textit{Sealer}$_2$ then compiles the modified source code and executes the returned program $\mathsf{P_{\mathcal{M}}}$. Honest sealers run $\mathsf{P_{\mathcal{O}}}$ while malicious \textit{Sealer}$_2$ operates $\mathsf{P_{\mathcal{M}}}$. We observe the acceptance rate of each sealer's blocks by outputting the block sealer's address accompanied by the \textit{diff} field (using methods in Appendix A). If the miner address of each block turns out to be the same, namely $addr_2$, \textit{Sealer}$_2$ achieves the goal of frontrunning the original in-turn sealer and producing blocks that outpace others. In this case, our attack is considered to be successful.
Figure~\ref{fig-attack} demonstrates the detailed procedures of verification and the attack.

\begin{figure}[hpt!]
    \centering
    \includegraphics[width=0.9\linewidth]{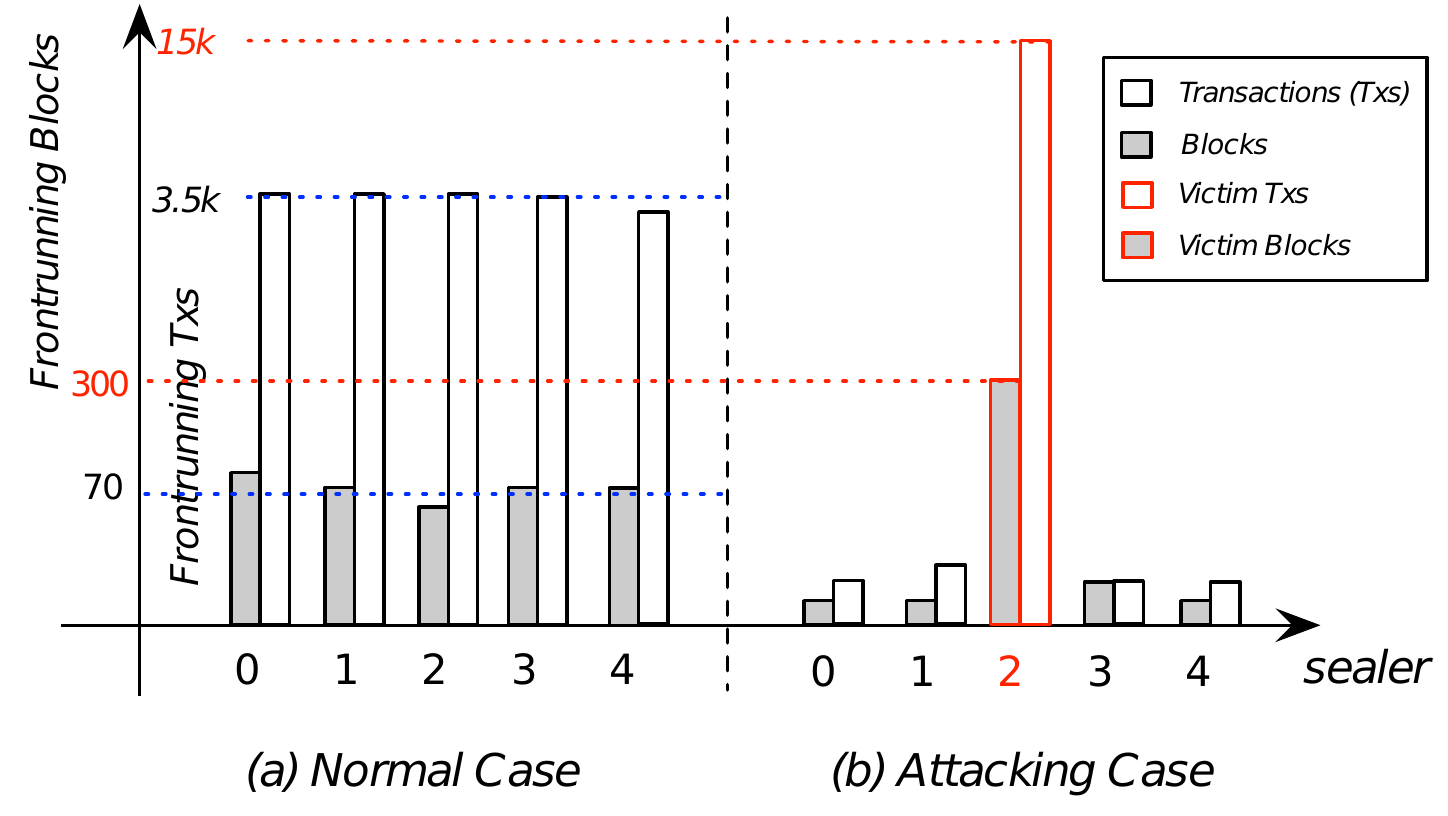}
   \caption{Experimental Results of Created Blocks/Transactions}
    \label{fig-result}
\end{figure}

\smallskip
\noindent\textbf{Experimental Result.}
Figure~\ref{fig-result} demonstrates the experimental results of both block and transaction creation.
In the first experiment, all sealers generate $360$ blocks within $30$ minutes in total, and each of them generates $70$ blocks approximately. This indicates that these sealers generate blocks with almost equal possibilities. Meanwhile, we observe that the transactions are equally distributed in each block (50 TXs/block) as well. On the contrary, in the second experiment, most of the blocks (over $300$ blocks) are generated by \textit{Sealer}$_2$ ($\mathsf{0xa5a...}$, see the screenshots in Appendix A), and the transactions in \textit{Sealer}$_2$'s blocks are over 15K in total. The results demonstrate that a malicious sealer running $\mathsf{P_{\mathcal{M}}}$ can successfully frontrun both blocks and transactions.

\section{Attacking Analysis}
\label{sec-seurity}
In this section, we dive into the root reason which makes such attacks feasible. We then analyze the attacking impacts for the blockchain ecosystem.

\smallskip
\noindent\textbf{Root Reason.}
The original Clique protocol adopted by the implementation of Ethereum is equipped with three critical verification conditions upon block receipt. First, the block is considered invalid if its sealer has signed a block in recent $\left\lfloor\frac{N}{2}\right\rfloor+1$ turns, which means that only $\left\lceil\frac{N}{2}\right\rceil-1$ can propose a block in each round (cf. line \ref{clique:line:verrecent} Algorithm \ref{alg-proposeclique}). Second, the valid weight of a block is restricted to only $1$ or $2$ (line \ref{clique:line:verdiff}). Thirdly, the validity of a block weight needs to be verified corresponding to the sealer's identity, that is, whether the sealer holds a valid leader position when it penetrates a block with a weight equal to $2$ (line \ref{clique:line:verinturn}). The main weakness of Clique applied in our target network is its exceptional logic in difficulty validity check. By the design, Clique applies the rotation schema to ensure that sealers take turns to sign a block. However, without the $sign\_recently$ check and $inturn\_check$ verification, a malicious sealer can easily break this logic of block generation: \textit{the sealer can simply propose their blocks continuously with a weight (equiv. diff) equal to $2$, and thus holds priority in block acceptance with only little modifications to the local client}. In addition, there is a latency in the proposing procedure for honest in-turn sealers (line \ref{clique:line:recent}). However, a malicious sealer can also shorten the latency in order to achieve a higher success rate.

\smallskip
\noindent\textbf{Attacking Impact.}
The frontrunning block attack brings two consequences.  Firstly, it is harmful to other honest sealers. With constant attacks from malicious sealers, the mining opportunity for honest sealers may reduce significantly. In this case, the honest sealers may lose mining rewards that they should have obtained. Secondly, such attacks damage the interests of users. Since all the transactions are visible to the sealers, malicious sealers with advanced knowledge can detect a profitable trade and replace these users' transactions with their own ones by performing frontrunning block attacks. What's worse, malicious sealers may arbitrarily drop users' transactions, directly harming the users' interests. The attack does not harm the target blockchain's \textit{consistency} or \textit{liveness}. However, it undermines the \textit{fairness} of block orders in practice. Notably, HPB maintains its decentralized trading platform Hpdex~\cite{hpdex}, which can be vulnerable to the attack. 

Table~\ref{tab-properties} compares the properties broken by different attacks, while Table~\ref{tab-cmp} shows the main differences among impacts of several attacks. Involving continuous controls over block issuance, our attack may cover a relatively wider range of victims compared to other known attacks. As sandwich attacks are achieved by frontrunning and backrunning a pair of profitable transactions at the same time, an evil sealer can easily commit such attacks on top of our attacking situations.

\begin{table*}[ht]
\caption{Comparison on security properties}\label{tab-properties}
\centering
\begin{threeparttable}
\resizebox{\linewidth}{!}{
\begin{tabular}{lccccr}
\toprule
\multicolumn{1}{c}{\textbf{\textit{Attack}}}   &
\multicolumn{1}{c}{\textbf{\textit{Assumption}}}  & \multicolumn{1}{c}{\textbf{\textit{Consistency}}}   &  \multicolumn{1}{c}{\textbf{\textit{Liveness}}}   & \multicolumn{1}{c}{\textbf{\textit{Fairness}}}   & \multicolumn{1}{c}{\textbf{\textit{Note}}}    \\ \midrule
This work   & honest sealers $\textgreater \frac{N}{2}^1$    &\Checkmark   & \Checkmark & \XSolidBrush  & block-level attack against in-turn sealers \\
PoA attack Type-I \cite{wang2022poa}  &  honest sealers $\textgreater \frac{N}{2}$, attacker honest but \textit{rational}$^2$ & \Checkmark   & \Checkmark & \XSolidBrush & transaction-level \\
PoA attack Type-II \cite{wang2022poa} &  honest sealers $\textgreater \frac{N}{2}$, attacker honest but \textit{rational} & \Checkmark   & \Checkmark & \XSolidBrush & block-level attack against edge-turn sealers  \\
Sandwiching$^3$ \cite{zhou2021high}  & depend on referenced consensus protocol &\Checkmark   & \Checkmark & \XSolidBrush & transaction-level    \\
Clone attack \cite{vincent20poa} &  honest sealers $\textgreater \frac{N}{2}$ & \XSolidBrush   & \XSolidBrush   & N/A & block-level attack for double-spending   \\
51\% attack~\cite{51attack}      & more than $50\%$ computational power controlled by a miner group  & \XSolidBrush   & \XSolidBrush   & N/A & double-spending   \\ \bottomrule
\end{tabular}
}
\begin{tablenotes}
       \footnotesize
       \item[1] The assumption means the malicious tolerance in PoA protocols is $t\textless\frac{N}{2}$.
       \item[2] \textit{Rational}: Seeking to maximize their profits within the scope of code principles
       \item[3] Sandwiching is presented as a practical instance of profitable transaction order manipulation, rather than attacks at the consensus level.
       \item[4] \XSolidBrush  indicates the property is vulnerable under current attack, while \Checkmark means the opposite.
\end{tablenotes}
\end{threeparttable}
\end{table*}

\section{Countermeasures}
\label{sec-countermeasures}
The verification procedure of original Clique protocol is equipped with the three critical conditions (cf. lines \ref{clique:line:verrecent}-\ref{clique:line:verinturn} Algorithm \ref{alg-fixpoint}). However, the derivative protocols may underestimate their importance. In the case of our target blockchain network HPB, we find the logic error in conditional expressions of the code. The error makes the system lack verification conditions related to both sealer's validity and the correspondence between a block's weight and its sealer's identity. 
The countermeasure is straightforward. The code logic error makes HPB lack two crucial verification checks, and this absence further results in vulnerabilities of HPB protocols. Therefore, fixing the mode logic error can directly plug the loophole. It is worth mentioning that official HPB team has already accepted our proposal (PR report~\cite{go-hpb-pr}). We suggest that future projects based on Clique could cautiously and comprehensively assess their security guarantees.

Algorithm~\ref{alg-fixpoint} highlights the fix points to the corresponding blockchain code. Since the clients can be compiled locally and the block proposing procedure is conducted independently, malicious sealers can easily falsify the proposing related code and manipulate the proposing parameters. As a consequence, the most effective way to prevent such manipulation is to strengthen the verification procedure, which is deployed on every sealer node, since it is difficult to compromise most nodes at the same time. As for flawed clique-based networks and future projects, rechecking the code logic and letting verification functions remain effective is essential, including \textit{sign\_recently, difficulty\_check} and \textit{inturn\_check}.

\begin{algorithm}
\caption{Block Verify in Clique}
\label{alg-fixpoint}
\begin{algorithmic}[1]
\Procedure{Verify}{$sealer_i$}
\Comment{\textcolor{blue}{verify a block}}
\State $\neg sign\_recently(sealer_i,block.number)$ \Comment{\textcolor{green}{fix point}} \label{clique:line:verrecent}
\State $\neg difficulty\_check(sealer_i,block.weight)$  \label{clique:line:verdiff}
\State $\neg inturn\_check(sealer_i,block.number)$ \Comment{\textcolor{green}{fix point}} \label{clique:line:verinturn}
\State $ \textbf{return}$ $\mathsf{true}$ \Comment{\textcolor{blue}{block is valid}}
\EndProcedure

\end{algorithmic}
\end{algorithm}

\section{Related Work} 
\label{sec-rw}

In this section, we sort out the security vulnerabilities that have arisen in PoA, including related research and various attacks proposed in the recent decade. 

\smallskip
\noindent\textbf{Insecurity of PoA.} Angelis \textit{et al.} \cite{de2018pbft} presented the first clear PoA protocol and clarified its impossibility under the CAP theorem~\cite{gilbert2002brewer}. PoA is inadequate in achieving consistency when prioritising another two properties. Vincent \textit{et al.} \cite{vincent20poa} proposed the cloning attack to break the consistency guarantee. A malicious sealer is granted abilities to clone itself into another instance and isolate two equivalent branches to influence the chain selection. A conflicting transaction, in this way, can be double-spent or discarded from the canonical chain. Toyoda \textit{et al.} \cite{toyoda2020function} dived into a function-level analysis by introducing a profiling tool \textit{pprof}. They detected the targeted functions that may decrease performance. Liu \textit{et al.} \cite{liu2019mdp} provided a quantitative analysis towards PoA by using Markov Decision Process. They launched a related attack against the VeChainThor's~\cite{vechain} PoA implementation. This work differs from the previous attacks in focusing on internal programming logic vulnerabilities.

\smallskip
\noindent\textbf{Fairness in PoA.}
 Besides the essential properties of a consensus protocol, \textit{consistency} and \textit{liveness}, Kelkar \textit{et al.}~\cite{kelkar2020order}~\cite{kelkar2021order} identified the third consensus property, \textit{transaction order-fairness}. Since the manipulation of transaction orders with non-public knowledge~\cite{markham1988front} may subvert the profitability or validity of target transactions, the term \textit{fairness} emphasizes transactions should be output upon the order of receipt to ensure the network users stay in a fair queue. Emerging MEV attacks against the blockchain fairness like~\cite{daian2020flash}\cite{qin2021quantifying} mainly focus on PoW consensus protocols. Stathakopoulou~\textit{et al.}~\cite{stathakopoulou2021adding} presented a hardware solution to prevent damage on \textit{fairness}. Recently, Wang \textit{et al.}~\cite{wang2022poa} exploited the \textit{fairness} attacks on PoA protocols. They pointed out underlying ways to manipulate transaction orders in PoA both on transaction-level and block-level. As for the block-level attack, they exploited the potentials of frontrunning edge-turn blocks (equiv. breaking fairness of second-priority blocks) when the in-turn sealer fails for reasons such as network congestion or being attacked. In contrast, our work focuses on the attack against the in-turn sealer (breaking fairness of first-priority blocks). We proposed an approach to compete for the block issuance right of in-turn sealers in each round with only one single evil sealer.

\begin{table}[ht]
\centering
\caption{Comparison among various attacks}
\label{tab-cmp}
\begin{tabular}{lcccr}
\toprule
\multicolumn{1}{c}{\textbf{\textit{Attack}}}      & \multicolumn{1}{c}{\textbf{\textit{Attacker}}}  & 
\multicolumn{1}{c}{\textbf{\textit{Miner Damage}}}  & 
\multicolumn{1}{c}{\textbf{\textit{Trader Damage}}}  \\ \midrule
This work  & Miner        & \Checkmark  & \Checkmark \\
PoA Type-I/II \cite{wang2022poa}  & Miner      & \Checkmark  & \Checkmark \\
Sandwiching~\cite{zhou2021high}  & Trader  &    \XSolidBrush  & \Checkmark \\
Clone attack~\cite{vincent20poa}   & Miner  &   \XSolidBrush  & \Checkmark \\
\bottomrule
\end{tabular}
\end{table}

\section{Conclusion}
\label{sec-conclusion}

Proof of Authority is one of the most prevailing consensus mechanisms in state-of-the-art permissioned blockchain systems. In this paper, we first explored the security issue of a Clique-based protocol used in the real-world projects. Then we focused on the block-level frontrunning attack by breaking the proper order of block issuance. We further launched the attack on HPB private chain, simulating strategies by falsifying priority parameters and maintaining an advantageous position during the sealer rotation.
The attack impeded consensus stability and caused monetary loss for both normal users and honest sealers. Results from our experiments proved the effectiveness of our frontrunning attack. Moreover, we provided our recommendations and disclosed the report to the official HPB team (cf. \cite{go-hpb-pr}). We hope that other Clique variants can draw upon this knowledge and avoid similar design mistakes. 

\section*{Acknowledgement}
Xinrui Zhang, Rujia Li and Qi Wang are partially supported by the Shenzhen Fundamental Research Programs under Grant No.20200925154814002. Lastly, we thank the HPB development team, especially for their open and helpful discussion.

\normalem
\bibliographystyle{IEEEtran}
\bibliography{bib}

\section*{Appendix A}

\begin{figure}[h!]
\subfigure[Normal Case]{
\begin{minipage}[b]{\linewidth}
\centering
\includegraphics[width=0.95\linewidth]{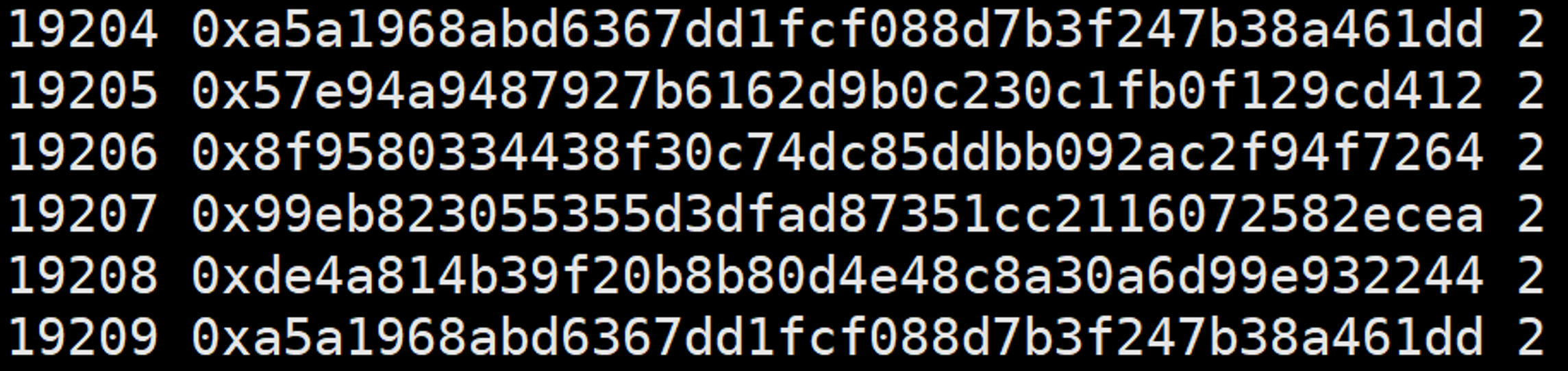}
\label{fig-screenshot-a}
\end{minipage}
}
\subfigure[Attacking Case]{
\begin{minipage}[b]{\linewidth}
\centering
\includegraphics[width=0.95\linewidth]{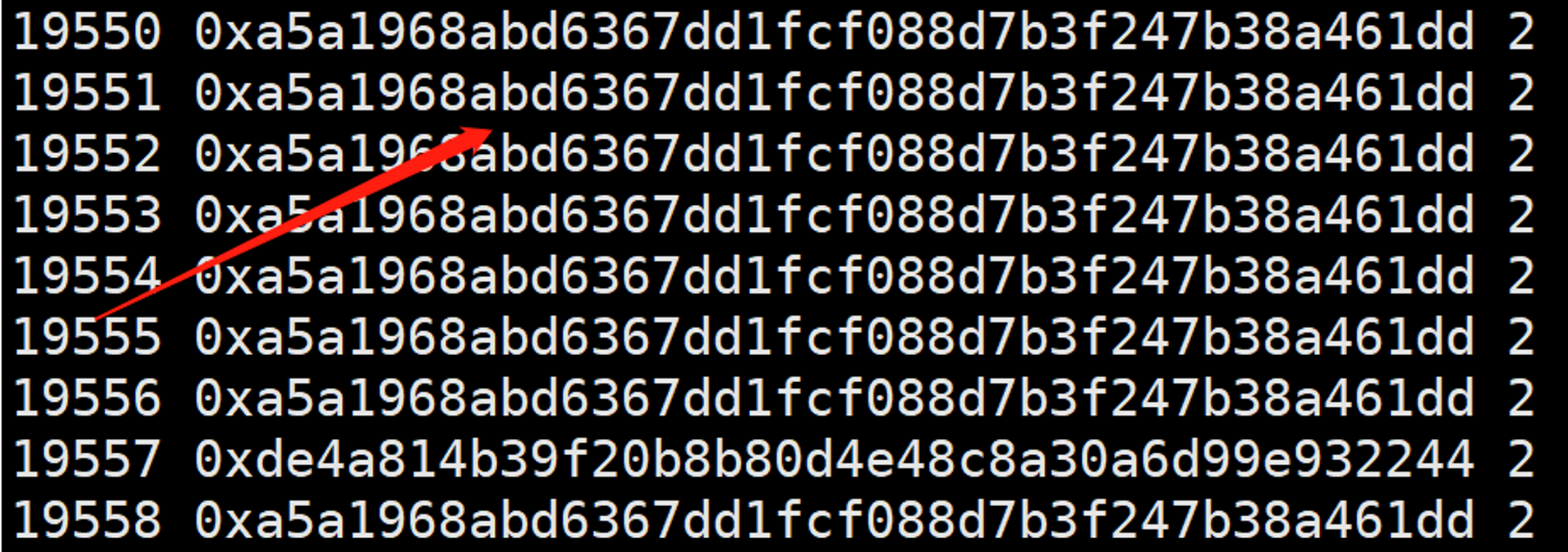}
\label{fig-screenshot-b}
\end{minipage}
}
\caption{Screenshots of Experimental Results}
\label{fig-screenshot}
\end{figure}

Figure \ref{fig-screenshot} shows a comparison between a normal case and an attacking case of running HPB instances. The first column on the left marks the serial number of each block. The middle column shows the addresses of in-turn sealers who have created current blocks. The right column represents the block weights (equiv. \textit{diff}). Figure \ref{fig-screenshot-a} depicts a normal case of a rotation cycle. We can observe that the blocks are signed by five different sealers in turns. The first five addresses ($addr_2$ -- $addr_5,addr_1$, sequentially) are not repeated while the sixth address ($addr_2$) rotates to the first one, corresponding to the rotation schema. In contrast, Figure \ref{fig-screenshot-b} shows an attacking case. Apparently, the addresses of nine consecutive blocks stay the same, indicating that a malicious sealer has successfully skipped the procedure of sealer rotation and sat in the leader position in most rounds. This out-of-turn sealer frontruns others to make profits by continuously producing profitable blocks by maintaining block weights (\textit{diff}) equal to 2.

We use the script (cf. Listing~\ref{lst:label}) to collect the above data. The output presents the block number, miner's address and block difficulty, sequentially.

\begin{lstlisting} [label={lst:label}, caption = Data collecting script]
#!/bin/bash
a=$1
for i in {0..199}
do
i=`expr $i + $a `
miner=`./ghpb attach rpc_addr  <<EOM
eth.getBlock($i).miner
EOM
`
difficulty=`./ghpb attach rpc_addr <<EOM
eth.getBlock($i).difficulty
EOM
`
echo $i ${miner:0-46:42} ${difficulty:0-4:2}
done
\end{lstlisting}

\section*{Appendix B. Notations}
\label{appendix:b}

We highlight several featured notations in the following Table~\ref{node} to clarify their usages in this work.

\begin{table}[H]
 \caption{Featured Notations}\label{tab-notation}
 \label{node}
  \centering
    \resizebox{\linewidth}{!}{  
    \begin{tabular}[t]{clr}
    \toprule
     \textit{\textbf{ Symbol }}  & \quad  \textit{\textbf{Item}} \quad  & \multicolumn{1}{c}{\textbf{ \textit{Functionalities}}}\\  \midrule
    \textit{diff} & Difficulty & the difficulty (weight value) of a block \\  
    \textit{delay time} & Delay time &  the delay time of current block issuance  \\
    $N$ & Committee size & the total number of sealers \\  
    $\mathsf{P_{\mathcal{O}}}$ & Original client & the normal (intact) client program of go-hpb \\  
    $\mathsf{P_{\mathcal{M}}}$ & Malicious client & the tampered client program of go-hpb  \\ 
    $S_\star$ & Sealer & refer to a sealer by index  \\ 
    $go$-$hpb$ & HPB client & the client program of HPB project \\ 
    $addr_\star$ & Sealer address & the address of a sealer \\
    \bottomrule
    \end{tabular}
    }
\end{table}

\section*{Appendix C}

Table.\ref{tab-projects} lists several mainstream commercial blockchain projects adopting Clique-based consensus algorithms. We present their market values collecting from public information (CoinMarketCap \cite{coinmarketcap}). The data demonstrate that Clique-related projects occupy a considerable market share, which emphasizes the importance of its protocol security.

\begin{table}[htb!] 
  \centering 
  \caption{Clique-based Projects}
  \label{tab-projects}
   \begin{threeparttable}
  \begin{tabular}[t]{lll}
    \toprule
      \textit{\textbf{Project}} & \textit{\textbf{Client}} & \textit{\textbf{Market Cap}} \\ \midrule
     
    Go-Ethereum  &  Clique    &  \$343,061,278,394 \\  
    Binance-chain  & Clique-variant   & \$17,426,222,620 \\  
     
    Polygon (MATIC) & Clique-forked   & \$9,866,753,531  \\ 
     
     Openethereum & Clique-forked   & \$343,061,278,394  \\  
     
     PoA network  &  Clique-variant    & \$23,723,240  \\ 
     
      HPB (\textit{\textbf{This work}})  & Clique-variant & \$3,058,901 \\
     
      Ethereum Classic & Clique-forked   & \$4,161,701,776  \\  
     
      ConsenSys & Clique-forked   &  -   \\  
     
     GoChain & Clique-forked    & \$24,702,737  \\
     
      Daisy  & Clique-forked  & -  \\
    
     Olecoin  & Clique-forked   & - \\ 
     
     EEX   & Clique-forked  & - \\ 
    
     AplaProject  & Clique-forked   & - \\ 
     
     Tomochain & Clique-forked  & \$114,756,610 \\ 
    \bottomrule 
  \end{tabular}
    \begin{tablenotes}
       \footnotesize
       \item[1] The \textit{variant} means a project makes modifications, while \textit{forked} means directly uses the source code from Clique.
       \item[2] Data accessed in April, 2022.
     \end{tablenotes}
  \end{threeparttable}     
\end{table}

\end{document}